\pdfoutput=1
\documentclass[notitlepage,aps,nofootinbib,
letterpaper,prl,twocolumn,amsmath,
amssymb,amsfonts,long,superscriptaddress,preprintnumbers]{revtex4-1}
\usepackage{amsmath,amssymb,amsthm,amsxtra,overpic,bbm,bm,epsfig,subfigure}
\usepackage{simplewick}
\usepackage[dvipsnames]{xcolor}
\usepackage{epstopdf}
\usepackage{epsfig}
\usepackage{comment}
\usepackage{graphicx,color,soul}
\usepackage{multirow}
\usepackage{verbatim}
\usepackage{enumitem}
\usepackage{hyperref}
\hypersetup{
	colorlinks=true,
	linkcolor=black,
	filecolor=blue,      
	urlcolor=blue,
	citecolor=blue,
}
\usepackage{lipsum}

\newcommand{\prlsection}[2]{{\it\textbf{#1}{#2}}---}
\makeatletter
\newcommand*{\balancecolsandclearpage}{%
	\close@column@grid
	\cleardoublepage
	\twocolumngrid
}
\makeatother

\begin{document}

\title{Solving the Hubble tension without spoiling Big Bang Nucleosynthesis 
}

\author{Guo-yuan Huang}
\email{guoyuan.huang@mpi-hd.mpg.de} 
\author{Werner Rodejohann}
\email{werner.rodejohann@mpi-hd.mpg.de} 
\affiliation{Max-Planck-Institut f\"ur Kernphysik, Postfach
	103980, D-69029 Heidelberg, Germany}

\date{\today}

\begin{abstract}
\noindent
The Hubble parameter inferred from  cosmic microwave background observations is consistently lower than that from local measurements, which could hint towards new physics. Solutions to the Hubble tension typically require a sizable amount of extra radiation $\Delta N^{}_{\rm eff}$ during recombination. However, the amount of $\Delta N^{}_{\rm eff}$ in the early Universe is unavoidably constrained by Big Bang Nucleosynthesis (BBN), which causes problems for such solutions.  
We present a possibility to evade this problem by introducing  
neutrino self-interactions via a simple Majoron-like coupling. The scalar is slightly heavier than $1~{\rm MeV}$ and allowed to be fully thermalized throughout the BBN era. The rise of neutrino temperature due to the entropy transfer via $\phi \to \nu\overline{\nu}$ reactions compensates the effect of a large $\Delta N^{}_{\rm eff}$ on BBN. Values of $\Delta N^{}_{\rm eff}$ as large as $0.7$ are in this case compatible with BBN. We perform a fit to the parameter space of the model. 
\end{abstract}

\preprint{}

\maketitle

\prlsection{Introduction}{.}%
The Hubble parameter inferred from the Planck observations of the cosmic microwave background (CMB), $H^{}_{0} = 67.4 \pm 0.5~{\rm km/s/Mpc}$~\cite{Aghanim:2018eyx}, is in  tension with that of  local measurements at low red-shifts. To be specific, the result from the Hubble Space Telescope (HST) by observing the Milky Way Cepheids is $H^{}_{0} = 74.03 \pm 1.42~{\rm km/s/Mpc}$~\cite{Riess:2019cxk}, which exceeds the value of Planck experiment by a $4.4 \sigma$ significance. Combining the HST result with a later independent determination~\cite{Freedman:2019jwv} yields a lower value of $H^{}_{0} = 72.26 \pm 1.19~{\rm km/s/Mpc}$, but the tension still persists at $3.7\sigma$ level.

The tension for the Hubble parameter could suggest the existence of new physics beyond the Standard Model or beyond the $\Lambda$CDM framework \cite{DiValentino:2020zio}.  
There is  a strong positive correlation between $H^{}_{0}$ and an extra radiation,  $\Delta N^{}_{\rm eff} = N^{}_{\rm eff} -3.046$, in the early Universe. Hence, by increasing $N^{}_{\rm eff}$ during recombination one can lift the Hubble parameter. However, increasing $N^{}_{\rm eff}$ delays matter-radiation equality and modifies the CMB power spectrum. This, in turn, can be compensated by introducing non-standard neutrino self-interactions during  recombination  \cite{Lancaster:2017ksf,Kreisch:2019yzn}. Thus, a successful particle physics model to explain the Hubble tension needs to provide a significant amount of $\Delta N^{}_{\rm eff}$ in the early Universe, as well as ``secret" neutrino interactions. 
In the original fits with Planck 2015 data, two modes with self-interacting neutrinos of the form $G^{}_{\rm eff} \overline{\nu}\nu \overline{\nu}\nu$ are identified, which are given in Table~\ref{table:modes}.
\begin{table}[b!] 
	\centering
	\begin{tabular}{|c| c  c  c|} 
		\hline
		Parameter & ${\rm log}^{}_{10}( G^{}_{\rm eff} \cdot {\rm MeV^2} )$ & $\Delta N^{}_{\rm eff}$ & $\eta^{}_{10}$ \\ 
		\hline
		SI$\nu$ & $-1.35^{+0.12}_{-0.07}$ & $1.02 \pm 0.29$ & $6.151^{+0.079}_{-0.090}$ \\ 
		MI$\nu$ & $-3.90^{+1.00}_{-0.93}$ & $0.79 \pm 0.28$ & $6.253 \pm 0.082$ \\ 
		\hline
	\end{tabular}
	\vspace{0.4cm}
	\caption{Central values and $1\sigma$ ranges of two modes in the fit of Planck 2015 data~\cite{Kreisch:2019yzn}. SI$\nu$ (MI$\nu$) stands for the strongly (moderately) interacting neutrino mode; $\eta^{}_{10} \equiv \eta^{}_{\rm b} \times 10^{10}$ represents the baryon-to-photon ratio. These results are updated with the Planck 2018 data in Refs.~\cite{Choudhury:2020tka,Brinckmann:2020bcn,Das:2020xke,Mazumdar:2020ibx}.}
	\label{table:modes}
\end{table}
The mode with strongly interacting neutrinos (SI$\nu$) is basically excluded by various terrestrial  experiments~\cite{Blinov:2019gcj,Lyu:2020lps,Brdar:2020nbj,Deppisch:2020sqh,Brune:2018sab}, so we shall confine ourselves to moderately interacting neutrinos (MI$\nu$). 

Among many attempts~\cite{Archidiacono:2020yey,Kelly:2020aks,He:2020zns,Berbig:2020wve,Seto:2021xua,Arias-Aragon:2020qip,Grohs:2020xxd,Escudero:2019gvw,Forastieri:2019cuf}, one of the simplest possibilities is the Majoron-like  interaction~\cite{Arias-Aragon:2020qip,Grohs:2020xxd,Escudero:2019gvw,Forastieri:2019cuf}\footnote{This type of coupling may be connected to the neutrino mass generation via the Majoron model~\cite{Chikashige:1980qk,Chikashige:1980ui,Gelmini:1980re,Choi:1991aa,Acker:1992eh,Georgi:1981pg,Schechter:1981cv}, where both scalar and pseudoscalar couplings exist after the spontaneous breaking of lepton number. For singlet Majorons it can be generated by mixing with heavy right-handed Majorana neutrinos in a gauge invariant UV completion. }
\begin{eqnarray} \label{eq:L}
	\mathcal{L} \supset  g^{}_{\alpha\beta} \phi \overline{\nu^{}_{\rm \alpha L}}   \nu^{\rm c}_{\rm \beta L} \;,
\end{eqnarray}
where $\alpha$ and $\beta$ run over flavors $e$, $\mu$ and $\tau$, and $g^{}_{\alpha \beta}$ are flavor-dependent coupling constants. 
After the neutrino temperature drops below $m^{}_{\phi}$, the interactions among neutrinos will be reduced to an effective coupling $G^{}_{\rm eff} = |g|^2_{}/m^{2}_{\phi}$. 
The flavor-specific couplings $g^{}_{ee}$ and $g^{}_{\mu\mu}$ are severely constrained by  laboratory searches~\cite{Blinov:2019gcj,Lyu:2020lps,Brdar:2020nbj,Deppisch:2020sqh,Brune:2018sab}, and we are only left with  $g^{}_{\tau\tau}$ to accommodate the MI$\nu$ mode during  recombination. 

In our model, the scalar particle $\phi$ with a mass $m^{}_{\phi}$ increases  $N^{}_{\rm eff}$. As long as the coupling in Eq.\ (\ref{eq:L}) is strong enough, $\phi$ will be in thermal equilibrium with the neutrino plasma, contributing to  extra radiation by $\Delta N^{}_{\rm eff} = 1/2 \cdot 8/7 \simeq 0.57$ for $m^{}_{\phi} \ll T^{}_{\nu}$, where $T^{}_{\nu}$ is the plasma temperature. Note that $\phi$ is in equilibrium before BBN as long as $g^{}_{\alpha \beta} \gtrsim 2.2 \times 10^{-10} ({\rm MeV}/m^{}_{\phi})$~\cite{Huang:2017egl}. 
However, as in many other models, this framework is put under pressure by the  primordial element abundances from Big Bang Nucleosynthesis (BBN)~\cite{Huang:2017egl,Schoneberg:2019wmt,Blinov:2019gcj,Venzor:2020ova}.
Incorporating the latest observations, BBN sets a strong constraint on the effective number of neutrino species~\cite{Pitrou:2018cgg}
\begin{eqnarray} \label{eq:Neff}
	N^{}_{\rm eff} = 2.88 \pm 0.27 \;.
\end{eqnarray}
This can be translated into a $2\sigma$ upper bound $\Delta N^{}_{\rm eff} < 0.42$, which severely limits the presence of extra radiation to solve the Hubble problem.

In this work, we explore a novel possibility that allows a large $\Delta N^{}_{\rm eff}$ surpassing the standard BBN constraint in Eq.~(\ref{eq:Neff}). In our Majoron-like model given in Eq.~(\ref{eq:L}), with $m^{}_{\phi} \gtrsim 1~{\rm MeV}$, the scalar particle can stay safely in thermal equilibrium throughout the epoch of BBN, in contrast to concerns in the  literature~\cite{Blinov:2019gcj,Dev:2019anc,Schoneberg:2019wmt}. Namely, since $m^{}_{\phi} \gtrsim 1~{\rm MeV}$, the neutrino temperature will increase with respect to the photon one due to $\phi \leftrightarrow \nu + \overline{\nu}$ reactions after the neutrinos have decoupled from the electromagnetic plasma at $T^{\rm dec}_{\nu} \sim 1~{\rm MeV}$. 
The rise in the neutrino temperature (increasing the neutron burning rate) will cancel the effect caused by a larger $N^{}_{\rm eff}$ (increasing the expansion rate), such that the final neutron-to-proton ratio $n/p$ remains almost the same as in the standard case. 

\begin{figure}[t!]
	\begin{center}
		\hspace{-1cm}
		\includegraphics[width=0.44\textwidth]{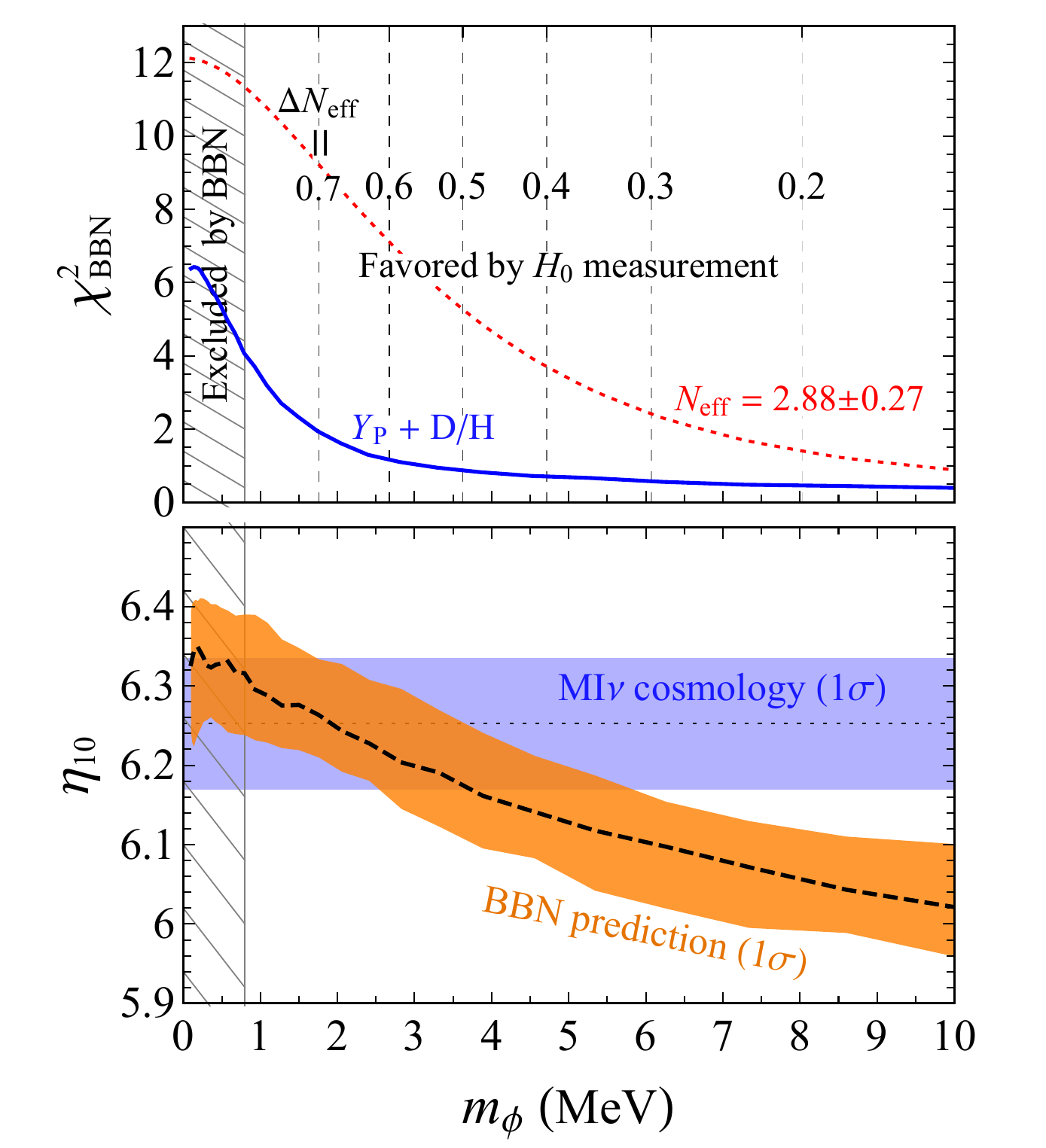}
	\end{center}
	\caption{The statistical significance of BBN $\chi^2_{\rm BBN}$ (upper panel) and baryon-to-photon ratio $\eta^{}_{10}$ (lower panel) as functions of the scalar mass $m^{}_{\phi}$, assuming $\phi$ is tightly coupled to all three active neutrinos throughout BBN. In the upper panel, the solid blue (or dotted red) curve shows $\chi^2_{\rm BBN}$ by fully simulating nucleosynthesis with the \texttt{AlterBBN} code~\cite{Arbey:2011nf,Arbey:2018zfh} (or adopting the usual bound $N^{}_{\rm eff} = 2.88 \pm 0.27$~\cite{Pitrou:2018cgg}). The dashed vertical lines stand for values of $\Delta N^{}_{\rm eff}$ for corresponding $m^{}_{\phi}$. A value  $\Delta N^{}_{\rm eff} \simeq 0.7$ with $m^{}_{\phi} \gtrsim 1~{\rm MeV}$ is permitted by BBN observations, $\chi^2_{\rm BBN} \simeq 2$, in contrast to the usual BBN bound $\Delta N^{}_{\rm eff} \lesssim 0.42$ at $2\sigma$ level~\cite{Pitrou:2018cgg}.
		In the lower panel, the yellow region signifies the $1\sigma$ allowed range of $\eta^{}_{10}$ predicted by helium-4 and deuterium abundances for different $m^{}_{\phi}$. The independent preferred range given by CMB fit~\cite{Kreisch:2019yzn} with moderately self-interacting neutrinos is shown in the horizontal blue band. It can be noticed that $m^{}_{\phi} \simeq 2~{\rm MeV} $ provides excellent fits to both BBN and CMB with MI$\nu$.}
	\label{fig:chi2}
\end{figure}

After a realistic BBN simulation is performed using Eq.~(\ref{eq:L}), we depict the chi-square function $\chi^2_{\rm BBN}$ as a function of the scalar mass $m^{}_{\phi}$ in the upper panel of Fig.~\ref{fig:chi2} (blue curve). This $\chi^2_{\rm BBN}$ includes the latest measurements of the helium-4 mass fraction ($Y^{}_{\rm P}$)~\cite{Aver:2015iza} and the deuterium abundance (${\rm D/H}$)~\cite{Cooke:2017cwo}, as well as various nuclear uncertainties.
The dotted red curve represents $\chi^2_{\rm BBN}$ obtained simply with Eq.~(\ref{eq:Neff}), i.e.\  without any scalar $\phi$ or rise in neutrino temperature, for the given $\Delta N^{}_{\rm eff}$. 
Parameters with $\chi^2_{\rm BBN} > 4$ are ruled out at $2\sigma$ level.
It can be observed that
a $\Delta N^{}_{\rm eff}$ value as large as $0.7$ is allowed for $m^{}_{\phi} =  1.8~{\rm MeV}$ without spoiling BBN, i.e., $\chi^2_{\rm BBN} \simeq 2$, in contrast to $\chi^2_{\rm BBN} \simeq 9$ using simply the $N_{\rm eff}$ value in  Eq.~(\ref{eq:Neff}). In the lower panel of Fig.~\ref{fig:chi2}, we also show the preferred baryon-to-photon ratio $\eta^{}_{10} \equiv \eta^{}_{\rm b} \times 10^{10}$ for each $m^{}_{\phi}$. Interestingly, the $1\sigma$ band around $m^{}_{\phi} = 2~{\rm MeV}$ matches very well with the independent determination of $\eta^{}_{10}$  from the CMB fit within the moderately self-interacting neutrino case \cite{Kreisch:2019yzn}. In contrast, the standard case, which can be roughly represented by $m^{}_{\phi}  = 10~{\rm MeV} \gg T^{\rm dec}_{\nu}$, disagrees with the MI$\nu$ value of $\eta^{}_{10}$ by nearly $2\sigma$.

In the following, we will illustrate the idea and results in more details.

\prlsection{Large extra radiation for BBN}{.}%
The improvements in the measurement of primordial element abundances and cross sections of nuclear reactions have made BBN an accurate test for  physics beyond the Standard Model~\cite{Zyla:2020zbs,Pitrou:2018cgg}.
The presence of extra radiation during the BBN era will accelerate the freeze-out of neutron-proton conversion, resulting in a larger helium-4 abundance than the prediction of standard theory. In addition, the abundance of deuterium is extremely sensitive to the baryon-to-proton ratio $\eta^{}_{\rm b}$, leaving BBN essentially parameter-free. 

\begin{figure}[t!]
	\begin{center}
		\hspace{-0.5cm}
		\includegraphics[width=0.47\textwidth]{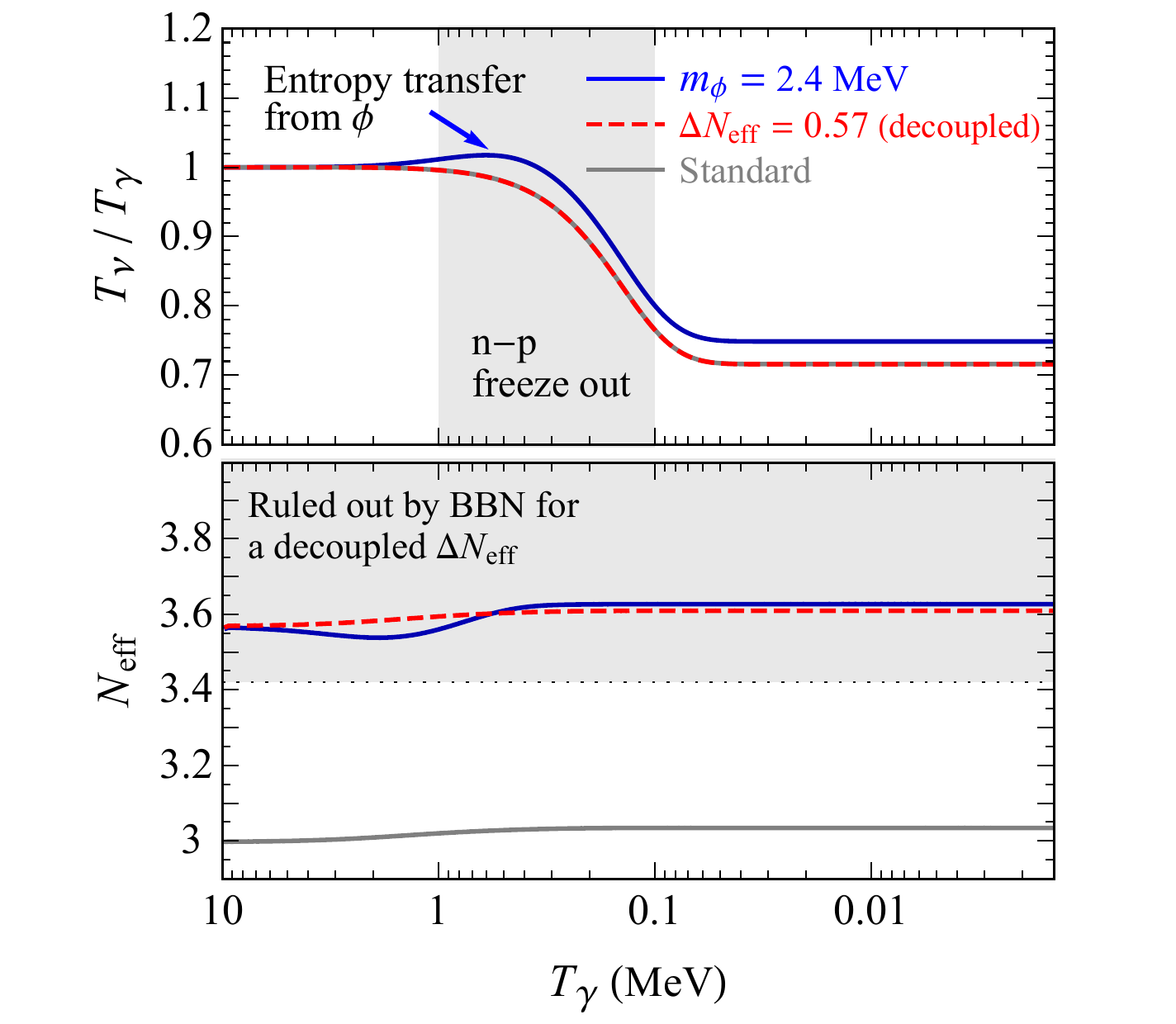}
	\end{center}
	\vspace{-0.3cm}
	\caption{The temperature ratio for neutrinos and photons $T^{}_{\nu}/T^{}_{\gamma}$ (upper panel) and $N^{}_{\rm eff}$ (lower panel) with respect to the photon plasma temperature. For all panels, the blue curves stand for the case of $m^{}_{\phi} = 2.4~{\rm MeV}$ assuming $\phi$ is in thermal equilibrium with neutrinos, while the red curves stand for the case with the decoupled $\Delta N^{}_{\rm eff}$. The gray curves signifies the standard case with neutrino-electron decoupling taken into account.}
	\label{fig:tem}
\end{figure}

The model-independent bounds as in Eq.~(\ref{eq:Neff}) are usually applicable to a ``decoupled'' $\Delta N^{}_{\rm eff}$, which is assumed to evolve separately from the Standard Model plasma. For the decoupled $\Delta N^{}_{\rm eff}$, the main effect is to change the expansion rate of the Universe, while leaving other ingredients untouched. 
However, this is not the case if $\phi$ tightly couples to neutrinos, such that  entropy exchange can take place between them. In our case, the argument based on $\Delta N^{}_{\rm eff}$ should be taken with caution, and we need to solve the primordial abundances.

Two steps are necessary to derive the light element abundances.
First, the background evolution of various species ($e^{\pm}$, $\gamma$, $\nu$ and $\phi$) needs to be calculated. Second, we integrate this into a BBN code to numerically simulate the synthesis of elements.
To calculate the evolution of background species, we solve the Boltzmann equations including the weak interactions between neutrinos and electrons, so the non-instantaneous decoupling of neutrinos is taken into account. More details can be found in the appendix.
We assume that all three generations of active neutrinos are in thermal equilibrium with $\phi$ before and during BBN, which holds for $g^{}_{\alpha \beta} \gtrsim 2.2 \times 10^{-10} ({\rm MeV}/m^{}_{\phi})$, such that one temperature $T^{}_{\nu}$ is adequate to capture the statistical property of the neutrino-$\phi$ plasma. This greatly boosts our computation without solving  discretized distribution functions.

In Fig.~\ref{fig:tem} we show the evolution of temperature ratio of neutrino to photon $T^{}_{\nu} / T^{}_{\gamma}$ (upper panel) and $N^{}_{\rm eff}$ (lower panel) as functions of the photon temperature $T^{}_{\gamma}$. Two beyond-standard-model scenarios are given: one with the tightly coupled Majoron-like scalar with mass $m^{}_{\phi} = 2.4~{\rm MeV}$ (blue curves), and one with the decoupled $\Delta N^{}_{\rm eff} = 0.57$ (red curves).
The standard case with only three active neutrinos is shown as gray curves.
Note from the lower panel that both scenarios are excluded if we simply adopt 
the $\Delta N^{}_{\rm eff}$ bound, i.e.\ if we disregard the effect of $\phi$-interactions on BBN.  
In the upper panel, for the case of $m^{}_{\phi} = 2.4~{\rm MeV}$, shortly after $T^{}_{\gamma} < m^{}_{\phi}$, the neutrino plasma receives entropy from the massive $\phi$, and its temperature is increased by $4.6\%$ compared to the standard value. In contrast, for the decoupled $\Delta N^{}_{\rm eff}$ scenario the ratio $T^{}_{\nu} / T^{}_{\gamma}$ is barely altered. 
Hence, different from the decoupled $\Delta N^{}_{\rm eff}$ scenario, there are two effects for the case $m^{}_{\phi} = 2.4~{\rm MeV}$: extra radiation $\Delta N^{}_{\rm eff}$ and  higher neutrino temperature $T^{}_{\nu}$.
If we assume the entropy from $\phi$ is completely transferred to neutrinos, the increased temperature can be calculated by using  entropy conservation. Namely,  $T^{\rm 0}_{\nu} = (g^{}_{*s} /g^{\rm 0}_{*s} )^{1/3} \, T^{}_{\nu} = (1+6.0\%)T^{}_{\nu}$, with $g^{}_{*s} \equiv 25/4$ and $g^{0}_{*s} = 21/4$ being the entropy degrees of freedom before and after $\phi$ decays, respectively.
In the realistic case, owing to the weak interactions between neutrinos and electrons, a small part of the entropy goes into the electron-photon plasma.

We now investigate these effects on the neutron-to-proton ratio $n/p$, which is the most important BBN quantity before the deuterium bottleneck at $T^{}_{\gamma} \simeq 0.078~{\rm MeV}$ is broken through.
For the neutron-proton conversion processes where neutrinos appear in the final state, e.g.\ $ {\rm p} + e^{-} \to {\rm n} + \nu^{}_{e} $,  the neutrino temperature $T^{}_{\nu}$ is relevant only through the Pauli blocking factor $1-f(p^{}_{\nu_e})$, which is insensitive to the small change of $T^{}_{\nu}$. Here, $f(p^{}_{\nu_e})$ stands for the Fermi-Dirac distribution function $f(p^{}_{\nu_e}) = 1/(1+\mathrm{e}^{p^{}_{\nu_e} / T^{}_{\nu}})$, where $p^{}_{\nu_e}$ is the momentum of $\nu^{}_{e}$ in the plasma.
Thus, we should be concerned about only two processes: ${\rm n} + \nu^{}_{e} \to {\rm p} + e^{-}$ and ${\rm p} + \overline{\nu^{}_{e}} \to {\rm n} + e^{+}$.
The rates are~\cite{Weinberg:2008zzc}
\begin{eqnarray} \label{eq:np1}
	\Gamma^{}_{{\rm n}  \nu^{}_{e}}  & = &  \frac{1}{\tau^{}_{\rm n} \lambda \, m^5_e}  \int^{\infty}_{0}  \mathrm{d} p^{}_{\nu_{e}} \sqrt{(p^{}_{\nu_e}+Q)^2 - m^2_e}  \notag\\
	&&	\frac{p^{}_{\nu_e} +Q}{1+ \mathrm{e}^{- (p^{}_{\nu_e} +Q)/T^{}_{\gamma}}} \cdot  \frac{p^{2}_{\nu_{e}}}{1+\mathrm{e}^{p^{}_{\nu_e}/T^{}_{\nu}}} \;, \\ \label{eq:np2}
	\Gamma^{}_{{\rm p} \overline{\nu^{}_{e}}}  & = &  \frac{1}{\tau^{}_{\rm n} \lambda \, m^5_e}  \int^{\infty}_{Q+m^{}_{e}}  \mathrm{d} p^{}_{\nu_{e}} \sqrt{(p^{}_{\nu_e}-Q)^2 - m^2_e}  \notag\\
&&	\frac{p^{}_{\nu_e} -Q}{1+ \mathrm{e}^{- (p^{}_{\nu_e} -Q)/T^{}_{\gamma}}} \cdot  \frac{p^{2}_{\nu_{e}}}{1+\mathrm{e}^{p^{}_{\nu_e}/T^{}_{\nu}}} \;,
\end{eqnarray}
where  $m^{}_{e}$ is the electron mass, $Q \equiv m^{}_{\rm n} - m^{}_{\rm p} \simeq 1.293~{\rm MeV}$, $\lambda \simeq 1.636$, and $\tau^{}_{n}$ the neutron lifetime. 
When $T^{}_{\nu} < Q$, the rate for ${\rm p} + \overline{\nu^{}_{e}} \to {\rm n} + e^{+}$ is suppressed by a Boltzmann factor $\mathrm{e}^{-Q/T^{}_{\nu}}$, i.e., only neutrinos with enough initial energy are kinematically allowed for the process. 
In contrast, the neutron-burning process ${\rm n} + \nu^{}_{e} \to {\rm p} + e^{-}$ can take place without  energy threshold.
Hence, after the decoupling of weak interactions at $T^{}_{\nu} \sim 1~{\rm MeV}$, a higher neutrino temperature compared to the standard case will result in a larger neutron burning rate. 
By ignoring the electron distribution function in the Pauli blocking factor and expanding the 
square-root by taking $m^{}_{e}/Q \simeq 0.15$ as a small quantity, the rate in Eq.~(\ref{eq:np1}) can be well approximated by
\begin{align} \label{eq:npa}
	\Gamma^{}_{{\rm n}  \nu^{}_{e}}   \simeq \frac{1}{\tau^{}_{\rm n} \lambda \, m^5_e}  &\left[ \frac{45 \zeta(5)}{2} T^5_{\nu} + \frac{7\pi^4}{60} Q T^4_{\nu} \right.\notag\\
	&	 \left. + \frac{3}{4}\left(2Q^2-m^{2}_{e} \right) \zeta(3) T^{3}_{\nu} \right] .
\end{align}
At low neutrino temperatures, the last term will dominate, i.e., $\Gamma_{{\rm n}  \nu^{}_{e} } \propto T^{3}_{\nu}$. 
Consequently, under the small perturbation of the neutrino temperature $\delta T^{}_{\nu}$,
the rate will be shifted by $\delta \Gamma/\Gamma_{{\rm n}  \nu^{}_{e} } = 3\, \delta T^{}_{\nu}/T^{}_{\nu}$. For the case of $m^{}_{\phi} = 2.4~{\rm MeV}$ in Fig.~\ref{fig:tem}, the relative temperature shift is about $\delta T^{}_{\nu}/T^{}_{\nu} = 4.6\%$, so we have $\delta \Gamma/\Gamma_{{\rm n}  \nu^{}_{e}} \simeq 13.8\%$.
During the temperature window $0.2~{\rm MeV} \lesssim T^{}_{\nu} \lesssim 1~{\rm MeV}$, 
the total neutron conversion rate $\Gamma^{\rm tot}_{\rm n}$ is mainly composed of two processes with similar rates, namely  ${\rm n} + \nu^{}_{e} \to {\rm p} + e^{-}$ and ${\rm n} + e^- \to {\rm p}  + \overline{\nu^{}_{e}}$, so we further have $\delta \Gamma/\Gamma^{\rm tot}_{\rm n} \simeq 6.9\%$. To conclude, for the case with $m^{}_{\phi} = 2.4~{\rm MeV}$, the change of neutrino temperature induced by the entropy transfer from $\phi$ will increase the total conversion rate from neutrons to protons  by almost $6.9\%$.

The above result will compensate the larger expansion rate caused by a positive $\Delta N^{}_{\rm eff}$. To see that, let us estimate more precisely the impact of $\Delta N^{}_{\rm eff}$.
Around the BBN era, the Hubble expansion rate is governed by
\begin{align} \label{eq:Hubble}
H \simeq \frac{1.66\sqrt{g^{}_{*}} T^2_{\gamma} }{ M_{\rm Pl}} \;,
\end{align}
where $g^{}_{*} = 5.5 +  7/4 \cdot N^{}_{\rm eff}$ stands for the relativistic degrees of freedom  before the annihilation of electrons, and $M_{\rm Pl} = 1.221 \times 10^{19}~{\rm GeV} $ for the Planck mass. Note that the time scales as $t \propto H^{-1}$. Hence, under a perturbation of $\Delta N^{}_{\rm eff}$, the amount of time over a certain temperature window (e.g.\ from $T^{}_{\gamma} = 1~{\rm MeV}$ to $T^{}_{\gamma} = 0.078~{\rm MeV}$) will be changed by $\delta t/t \simeq - 7/8 \cdot \Delta N^{}_{\rm eff} / g^{\rm std}_{*}$ with $g^{\rm std}_{*} = 10.75$ being the degrees of freedom with $N^{}_{\rm eff} = 3.046$.
For our case of $m^{}_{\phi} = 2.4~{\rm MeV}$, $\Delta N^{}_{\rm eff}$ is about  $0.6$, leading to $\delta t/t \simeq -4.9\%$. It is crucial that $\delta t/t$ is negative. 
The evolution of neutron number before $T^{}_{\gamma} = 0.1~{\rm MeV}$ is described by
\begin{align} \label{eq:n}
	\frac{\mathrm{d} n}{\mathrm{d} t} = - \Gamma^{\rm tot}_{\rm n} n + \Gamma^{\rm tot}_{\rm p}  p \;,
\end{align}
where $n$ and $p$ are the neutron and proton densities, respectively.
As has been mentioned before, the conversion rate from proton to neutron $ \Gamma^{\rm tot}_{\rm p} \simeq \Gamma^{\rm tot}_{\rm n} \mathrm{e}^{-Q/T^{}_{\nu}} $ is highly suppressed for $T^{}_{\nu} < Q \simeq 1.293~{\rm MeV}$.
After the decoupling of weak interactions at $T^{}_{\nu} \sim 1~{\rm MeV}$, the conversion from neutrons to protons will dominate the evolution of $n/p$.
Thus, the decreased neutron density in a small unit time window $t$ is given by $\Gamma^{\rm tot}_{\rm n}\, t \, n$, which is sensitive to both the perturbations of conversion rate and time (through the expansion rate).
As a result, the larger conversion rate with $\delta \Gamma/\Gamma^{\rm tot}_{\rm n} \simeq 6.9\%$ and the larger Hubble expansion rate with $\delta H/H \simeq 4.9\%$ (or $\delta t/t \simeq -4.9\%$) will  compensate each other.

\begin{figure}[t]
	\begin{center}
		\hspace{-1cm}
		\includegraphics[width=0.42\textwidth]{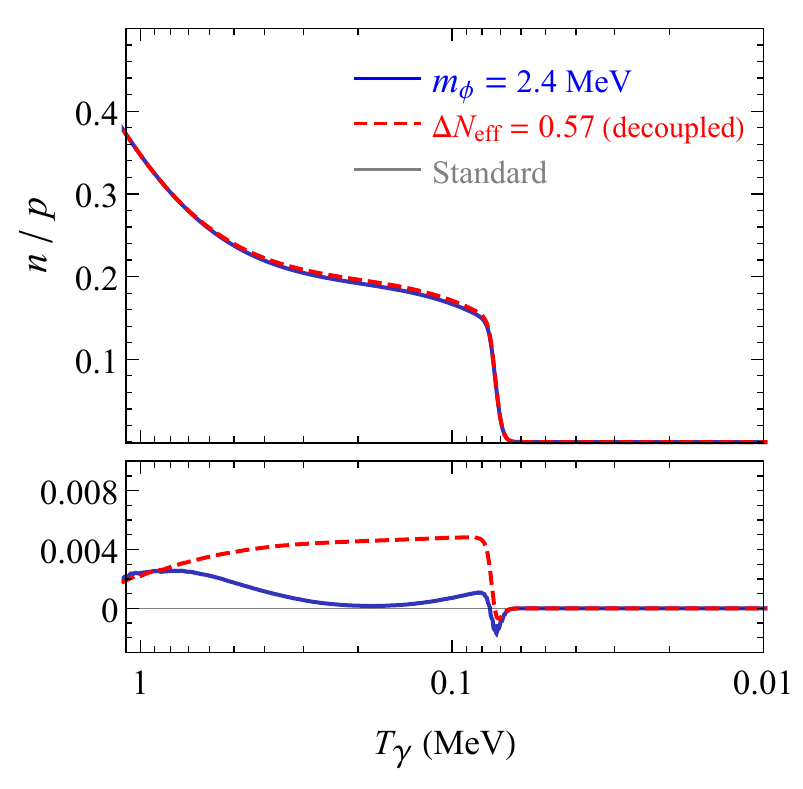}
	\end{center}
	\vspace{-0.3cm}
	\caption{The neutron-to-proton ratio shown as a function of the photon plasma temperature $T^{}_{\gamma}$ (upper panel). The lower panel gives the differences between the non-standard scenarios and the standard one. The convention is the same as for Fig.~\ref{eq:Neff}.}
	\label{fig:pnr}
\end{figure}
 
Having some analytical understanding, we adopt the \texttt{AlterBBN} code~\cite{Arbey:2011nf,Arbey:2018zfh} to calculate the light element abundances, incorporating the background quantities we solved before (see the appendix). We give in the upper panel of Fig.~\ref{fig:pnr} the evolution of $n/p$ with respect to the photon temperature. For illustration, the lower panel indicates the difference of two non-standard scenarios to the standard one, e.g., $n/p|^{}_{\phi} - n/p|^{}_{\rm std}$. One can clearly notice the different impacts of tightly-coupled $\phi$ and the decoupled $\Delta N^{}_{\rm eff}$. For the decoupled $\Delta N^{}_{\rm eff}$, $n/p$ takes a larger value than the standard one (by $\sim 0.005$) due to the higher expansion rate. This leads to a larger helium-4 abundance by $\delta Y^{}_{\rm p} \simeq 0.0076$, with $Y^{}_{\rm p} = 2 n/p (1+ n/p)$, which is significant given the error of the $Y^{}_{\rm p}$ measurement being about  $0.004$~\cite{Aver:2015iza}.
In contrast, for the case of a scalar $m^{}_{\phi} = 2.4~{\rm MeV}$, $n/p$ differs from the standard case only by $0.001$, resulting in a negligible change of helium-4 abundance $\delta Y^{}_{\rm p} \simeq 0.0015$. 
In this scenario, $n/p$ increases initially due to the larger expansion rate similar to the decoupled $\Delta N^{}_{\rm eff}$. Around $T^{}_{\gamma} \simeq 0.9~{\rm MeV}$, the increasing neutrino temperature starts to accelerate the burning of neutron, dragging $n/p$ back to the standard value. As a consequence, $Y^{}_{\rm p}$ is barely altered. 
This behavior agrees very well with the previous analytical observations.

\prlsection{Preferred parameter space}{.}%
Now we explore the preferred parameter space of the model by setting $m^{}_{\phi}$ and $g^{}_{\tau\tau}$ as free parameters, using BBN and other cosmological observations.
The primordial values of light element abundances can be inferred from the observation of  young astrophysical systems.
The mass fraction of helium-4 has been measured to be $Y^{}_{\rm p} = 0.2449 \pm 0.0040$ by observing the emission spectrum of low-metallicity compact blue galaxies~\cite{Aver:2015iza}.
In addition, the deuterium abundance with a much lower value  was derived by observing the absorption spectrum of Lyman-$\alpha$ forests above certain red-shifts.
The new recommended deuterium-to-hydrogen abundance ratio reads ${\rm D/H} = (2.527 \pm 0.030) \times 10^{-5}$~\cite{Cooke:2017cwo}.
On the other hand, the predicted value of $Y^{}_{\rm p}$ from BBN is dominated by the neutron-to-proton ratio $n/p$. A standard freeze-out value $n/p \simeq 1/7$ with $N^{}_{\rm eff} \simeq 3$ will give rise to $Y^{}_{\rm p} \simeq 0.25$. The synthesis of deuterium is more complex, depending on both $N^{}_{\rm eff}$ and $\eta^{}_{\rm b}$. The remarkable sensitivity of ${\rm D/H}$ to $\eta^{}_{\rm b}$ makes it an excellent baryon meter, especially with the recent update of deuterium-related nuclear rates~\cite{Coc:2015bhi}.
In the following, $Y^{}_{\rm p}$ and ${\rm D/H}$ will be used to constrain our model.

The mass of the scalar $\phi$ cannot be arbitrary in our scenario. If $m^{}_{\phi}$ is too large, the entropy will be mostly released before the neutrino decoupling epoch, and the resulting $\Delta N^{}_{\rm eff}$ is inadequate to explain the Hubble tension. 
On the other hand, if $m^{}_{\phi}$ is too small, there is not enough entropy transfer during the BBN era, and the BBN constraint on $\Delta N^{}_{\rm eff}$ cannot be evaded. 
This will confine the working range of $m^{}_{\phi}$, as seen in Fig.~\ref{fig:chi2}.

To fully explore the parameter space of scalar mass $m^{}_{\phi}$ and coupling constant $g^{}_{\tau\tau}$, we incorporate our results into a fit of CMB and large scale structure with self-interacting neutrinos. 
We note that the observational data of CMB and structure formation were initially used to derive bounds on the secret neutrino interactions~\cite{Hannestad:2004qu, Hannestad:2005ex, Bell:2005dr, Basboll:2008fx, Archidiacono:2013dua, Forastieri:2015paa, Cyr-Racine:2013jua, Oldengott:2014qra, Forastieri:2017oma,Oldengott:2017fhy}, but later a degeneracy was noticed between the effective neutrino coupling $G^{}_{\rm eff}$ and other cosmological parameters~\cite{Lancaster:2017ksf}, which can help to resolve the Hubble issue. With the Planck 2015 data, the Hubble tension can be firmly addressed by a large $\Delta N^{}_{\rm eff}$ along with self-interacting neutrinos~\cite{Kreisch:2019yzn}.
However, the fits based on the latest Planck 2018 data (specifically with the high-$l$ polarization data) show no clear preference for strongly interacting  neutrinos~\cite{Choudhury:2020tka,Brinckmann:2020bcn}. 
The results omitting the high-$l$ polarization data however remain similar to the analysis with Planck 2015 data. 
In either case, including the local measurement of $H^{}_{0}$ will always induce a preference for large $\Delta N^{}_{\rm eff}$ and non-vanishing $G^{}_{\rm eff}$, but the overall fit with the high-$l$ polarization data of Planck 2018 is poor.

\begin{figure}[t]
	\begin{center}
		\hspace{-0.5cm}
		\includegraphics[width=0.47\textwidth]{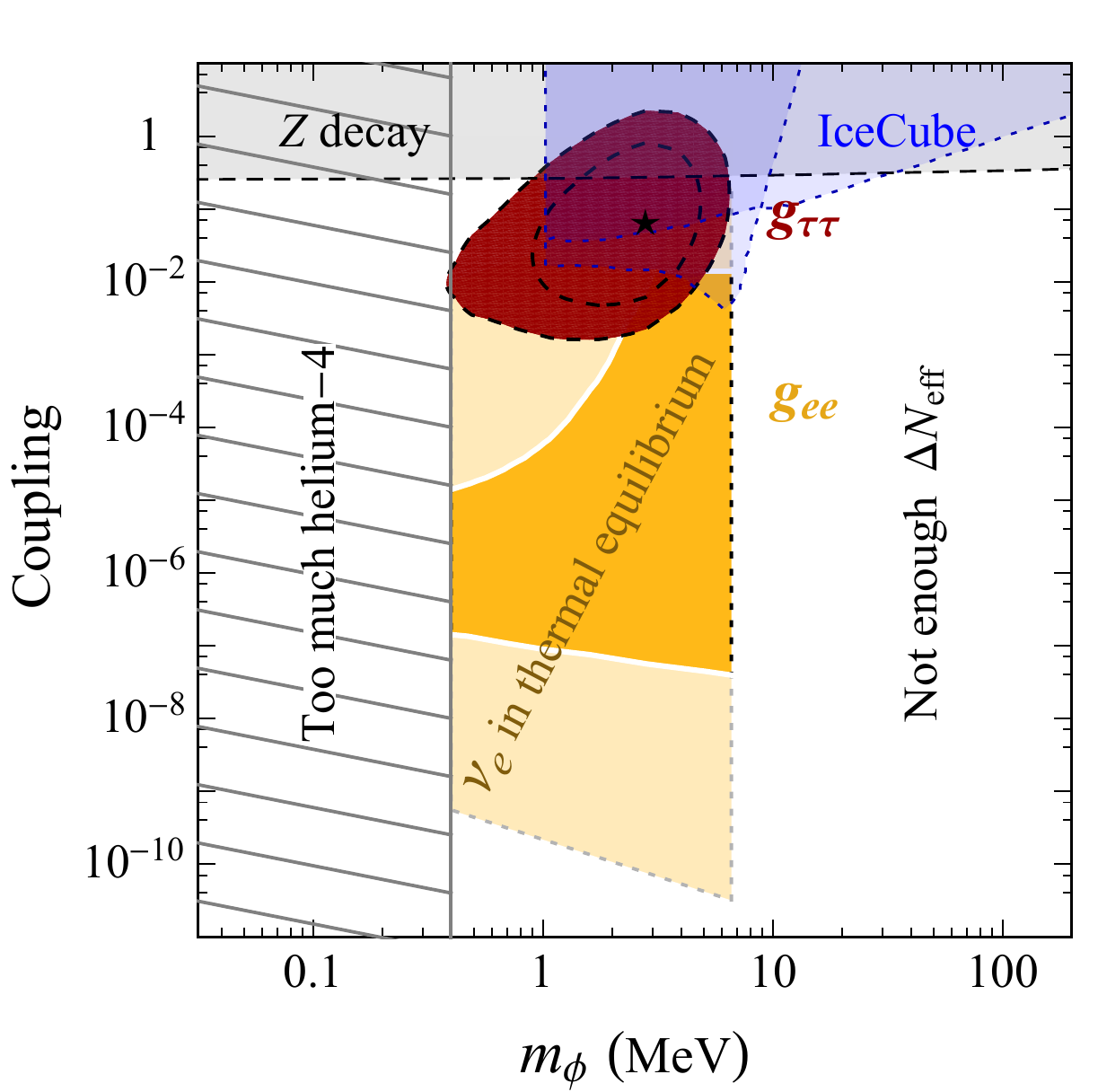}
	\end{center}
	\vspace{-0.3cm}
	\caption{The scalar coupling versus its mass $m^{}_{\phi}$. The red region is the $90\%$ preferred parameter space for $g^{}_{\tau\tau}$-$m^{}_{\phi}$ by taking BBN and fits of CMB and the local value of $H^{}_{0}$ into account~\cite{Mazumdar:2020ibx}. Note that the results of the moderately self-interacting neutrino mode have been used. Inside the red region, the dashed curve  surrounds the $1\sigma$ region, and the star in the middle marks the best-fit point. 
	The bound  on $g^{}_{\tau\tau}$ from $Z$ decays is shown in the gray band on the top~\cite{Brdar:2020nbj}.
	The required strength of $g^{}_{ee}$ to keep $\nu^{}_{e}$ and $\phi$ in equilibrium is shown in the yellow region, while various limits to $g^{}_{ee}$ are given as lighter yellow regions~\cite{Brune:2018sab,Blinov:2019gcj}. The IceCube limits on the universal couplings are recast as blue curves~\cite{Bustamante:2020mep}, which should be weakened for the flavor-specific coupling $g^{}_{\tau\tau}$.}
	\label{fig:gm}
\end{figure}
In order to be definite, we will adopt the results where only $\nu^{}_{\tau}$ moderately couples to $\phi$ during the recombination epoch. These include~\cite{Mazumdar:2020ibx}:
\begin{eqnarray} \label{eq:MInu}
&&{\rm log}^{}_{10}( G^{}_{\rm eff} \cdot {\rm MeV^2} )  = -3.2^{+ 1.3}_{-1.5} ,\;	
N^{}_{\rm eff} = 3.69^{+ 0.28}_{-0.33} , \; \notag\\ 
&&\eta^{}_{10} = 6.195\pm 0.099,
\end{eqnarray}
with the normal neutrino mass ordering.
The photon-to-baryon ratio $\eta^{}_{10}$ is converted from $\Omega^{}_{\rm b} h^2$ by using $\eta^{}_{10} = 274 \, \Omega^{}_{\rm b} h^2$~\cite{Steigman:2007xt}.
For each parameter choice of $m^{}_{\phi}$ and $g^{}_{\tau\tau}$, the total $\chi^2$ is constructed as a combination of the BBN one $\chi^{2}_{\rm BBN}$
and those fitted with the central value and symmetrized $1\sigma$ error in Eq.~(\ref{eq:MInu}).
The preferred region of parameter space is given in Fig.~\ref{fig:gm}.
The dark red region signifies the preferred parameter space for $g^{}_{\tau\tau}$ at $90\%$ confidence level (CL), inside which the dashed curve stands for the $68\%$ CL contour and the star represents the best-fit point, i.e., $m^{}_{\phi} = 2.8~{\rm MeV}$ and $g^{}_{\tau\tau} = 0.07$.

Laboratory and astrophysical searches set stringent upper limits on the secret neutrino interactions~\cite{Farzan:2018gtr,Lyu:2020lps,Brdar:2020nbj,Deppisch:2020sqh,Arcadi:2018xdd,Brune:2018sab,Bustamante:2020mep,Ng:2014pca,Ioka:2014kca,Ibe:2014pja,Kamada:2015era,Shoemaker:2015qul,DiFranzo:2015qea,Shalgar:2019rqe}, especially for the coupling strengths of $\nu^{}_{e}$ and $\nu^{}_{\mu}$ with the scalar. When it comes to the recombination epoch, to achieve  moderate self-interactions $G^{}_{\rm eff} \sim 10^{-3}~{\rm MeV}^{-2}$, we must have sizable $g^{}_{\tau\tau}$.
The bound on $g^{}_{\tau\tau}$ from $Z$ decays is shown as the gray band on the top~\cite{Brdar:2020nbj}.
On the other hand, to have a  higher neutron burning rate, $\nu^{}_{e}$ must stay in equilibrium with $\phi$ during the BBN era as we assumed in the previous discussion, which will impose a lower limit on the coupling constant $g^{}_{ee} \gtrsim 2.2 \times 10^{-10} ({\rm MeV}/m^{}_{\phi})$~\cite{Huang:2017egl}.
The required coupling strength for $g^{}_{ee}$ is depicted in the yellow region. The lighter yellow regions are excluded by neutrinoless double-beta decay experiments~\cite{Blum:2018ljv,Brune:2018sab} (top-left), $K$ decays~\cite{Blinov:2019gcj} (top-right) and supernova luminosity constraint~\cite{Brune:2018sab,Kolb:1987qy, Konoplich:1988mj, Farzan:2002wx, Zhou:2011rc,Heurtier:2016otg} (bottom), respectively.
The presence of large $g^{}_{\tau\tau}$ coupling will enhance the invisible decay rate of $Z$~\cite{Brdar:2020nbj}, which can be conveniently measured by the number of light neutrino species $N^{}_{\nu}$. Our best-fit case $m^{}_{\phi} = 2.8~{\rm MeV}$ and $g^{}_{\tau\tau} = 0.07$ predicts $N^{}_{\nu} = 3.0012$~\cite{Brdar:2020nbj}.

The region of our interest for $g^{}_{\tau\tau}$ may lead to a dip in the spectrum of ultra-high energy (UHE) neutrinos observed at IceCube by scattering off the relic neutrinos~\cite{Bustamante:2020mep,Ng:2014pca,Ioka:2014kca,Ibe:2014pja,Kamada:2015era,Shoemaker:2015qul,DiFranzo:2015qea}. If we assume the neutrino mass to be $m^{}_{\nu} = 0.1~{\rm eV}$, the resonant-scattering dip should be around $E^{}_{\nu} = m^2_{\phi} / (2 m^{}_{\nu}) \approx 78~{\rm TeV}$ for our best-fit case $m^{}_{\phi} = 2.8~{\rm MeV}$.
The absence of the dip at IceCube will place a constraint on our preferred parameter space~\cite{Bustamante:2020mep}, recast as dotted blue curves in Fig.~\ref{fig:gm}, which has covered part of ours $1\sigma$ parameter space. However, we need to mention that the actual constraints are subject to the neutrino mass spectrum, the neutrino flavor, the model of sources as well as initial spectrum of UHE neutrinos. For example, in some model where UHE neutrinos are generated from decays of dark matter in Milky Way, the constraints from diffuse spectrum do not apply anymore. The constraints in Ref.~\cite{Bustamante:2020mep} will also alter if a different spectrum index or flavor-specific coupling is taken.

\prlsection{Concluding remarks}{.}%
In this paper, we have explored the role of BBN for the Majoron-like scalar solution in light of the $H^{}_{0}$ tension. Note that this work is based on  scalar interactions of Majorana neutrinos, but similar or slightly modified considerations can also be made for other theories, e.g., complex scalar and vector interactions, and even Dirac neutrinos. 
By numerically solving the light element abundances, we find that a simple Majoron-like scalar with mass $\gtrsim {\rm MeV}$ can provide moderately self-interaction as well as large $\Delta N^{}_{\rm eff}$ during the recombination epoch to relieve the Hubble tension. The widely concerned BBN constraint on $\Delta N_{\rm eff}$ does not apply because it ignores the entropy transfer of a MeV-scale $\phi$ that heats up the neutrino bath. The extra radiation and the rise in the neutrino temperature are found to compensate each other, such that the large $\Delta N^{}_{\rm eff}$ is warranted throughout the BBN era, which should be very helpful to address the $H^{}_{0}$ tension.


\prlsection{Acknowledgement}{.}%
{\sl GYH would like to thank Kun-Feng Lyu for inspiring discussions. This work is supported by the Alexander von Humboldt Foundation.
}

\begin{appendix}
\section{Appendix}
In this appendix we explain how the evolution of the background is obtained in more details. 

In the assumption that three flavors of active neutrinos are all tightly coupled to $\phi$, only one temperature $T^{}_{\nu}$ is sufficient to describe the neutrino-$\phi$ plasma. The state of electron-photon plasma is represented by $T^{}_{\gamma}$. 
To  make the effect of expansion of the Universe explicit, it is convenient to introduce the following dimensionless quantities in the comoving frame:
\begin{align} \label{eqapp:cm}
x \equiv m a,~	q^{}_{i} \equiv p^{}_{i}\cdot a,~ z^{}_{i} \equiv T^{}_{i} \cdot a,~\widetilde{s}^{}_{i} = {s}^{}_{i} \cdot a^3 \;,
\end{align}
where $a$ is the dimensionful scale factor, $x$ the dimensionless scale factor with $m$ being an arbitrary mass scale (we set $m = 1~{\rm MeV}$), $q^{}_{i}$ the comoving momentum for species $i$, $z^{}_{i}$ the comoving temperature, and $\widetilde{s}^{}_{i}$ the comoving entropy density. If there is only one massless species in the Universe, $q^{}_{i}$, $z^{}_{i}$ and $\widetilde{s}^{}_{i}$ will be constant during the expansion of the Universe.

Taking account  the weak interactions,
the comoving entropy density transferred from the electron-photon plasma to the neutrino-$\phi$ one can be calculated with~\cite{Bernstein:1988bw,Grohs:2015tfy}
\begin{align} \label{eqapp:boltz}
H x	\frac{\mathrm{d}\widetilde{s}^{}_{\nu \phi}}{\mathrm{d}x} =  \sum_{\alpha}\int \frac{\mathrm{d}^3 q}{(2\pi)^3} C^{}_{\nu_\alpha}[f^{}_{e}, f^{}_{\nu}] \ln\left( \frac{f^{}_{\nu_\alpha}}{1-f^{}_{\nu_\alpha}}  \right) ,
\end{align}
where the collision terms $C^{}_{\nu_\alpha}[f^{}_{e}, f^{}_{\nu}]$ for six neutrino flavors $\nu_\alpha$ (including antineutrinos) have been widely calculated and are available in the literature, and see e.g., Refs.~\cite{Hannestad:1995rs,Dolgov:1997mb,Grohs:2015tfy}. When the collision terms are vanishing (i.e., no heating from the electron-photon plasma), the entropy in the neutrino-$\phi$ plasma is conserved. Also note that for the neutrino decoupling process, which is not a thermal equilibrium process, the total entropy $\widetilde{s}^{}_{\nu \phi} + \widetilde{s}^{}_{e \gamma}$ is not preserved in general.
The entropy density of the neutrino-$\phi$ plasma in terms of the neutrino comoving temperature reads~\cite{Bernstein:1988bw,Grohs:2015tfy}
\begin{align} \label{eqapp:snuphi}
 \widetilde{s}^{}_{\nu \phi} = - \sum_{i = \nu, \phi}\int \frac{\mathrm{d}^3 q}{(2\pi)^3} [ f^{}_{i} \ln f^{}_{i} \mp (1\pm f^{}_{i}) \ln(1 \pm f^{}_{i})  ] \;,
\end{align}
where the upper and lower signs apply to bosons and fermions, respectively. Here, the distribution functions for neutrinos and $\phi$ are $f^{}_{\nu} = 1/(1+\mathrm{e}^{q^{}_{\nu}/z^{}_{\nu}})$ and $f^{}_{\phi} = 1/(1-\mathrm{e}^{\sqrt{q^{2}_{\nu} + x^2 m^{2}_{\phi}/m^2}/z^{}_{\nu}})$, and the sum is over all neutrino species and the $\phi$ boson.
We use the following formula to establish the relation between the variation of entropy and that of the neutrino temperature ${\mathrm{d}z^{}_{\nu}}/{\mathrm{d}x}$:
\begin{align} \label{eqapp:st}
	\frac{\mathrm{d} \widetilde{s}^{}_{\nu \phi}}{\mathrm{d}x} = \frac{\partial \widetilde{s}^{}_{\nu \phi}}{\partial x} + \frac{\partial \widetilde{s}^{}_{\nu \phi}}{\partial z^{}_{\nu}} \cdot \frac{\mathrm{d}z^{}_{\nu}}{\mathrm{d}x} \;,
\end{align}
where ${\partial \widetilde{s}^{}_{\nu \phi}}/{\partial x}$ and ${\partial \widetilde{s}^{}_{\nu \phi}}/{\partial z^{}_{\nu}} $ can be straightforwardly obtained with Eq.~(\ref{eqapp:snuphi}). Some observations on Eq.~(\ref{eqapp:st}) are very helpful. If we assume there is no entropy transferred from other species, i.e., ${\mathrm{d} \widetilde{s}^{}_{\nu \phi}}/{\mathrm{d}x} = 0$, the variation of comoving temperature of neutrinos ${\mathrm{d}z^{}_{\nu}}/{\mathrm{d}x}$ will be proportional to ${\partial \widetilde{s}^{}_{\nu \phi}}/{\partial x}$. For $\widetilde{s}^{}_{\nu \phi}$, the only explicit dependence on $x$ is associated with the mass of $\phi$ in $f^{}_{\phi}$; therefore if $m^{}_{\phi} \ll T^{}_{\nu}$, ${\partial \widetilde{s}^{}_{\nu \phi}}/{\partial x}$ will be negligible and $T^{}_{\nu}$ simply follows the red-shift as the Universe expands. The function ${\partial \widetilde{s}^{}_{\nu \phi}}/{\partial x}$ roughly measures the entropy flow from $\phi$ to neutrinos.

On the other hand, the photon temperature can be derived by utilizing energy conservation $x \mathrm{d} \rho^{}_{\rm tot} / \mathrm{d} x = -3 (\rho^{}_{\rm tot} + P^{}_{\rm tot})$, namely
\begin{align} \label{eqapp:tgx}
x \frac{\mathrm{d}T^{}_{\gamma}}{\mathrm{d}x} =  \frac{-3 (\rho^{}_{\rm tot} + P^{}_{\rm tot}) 
- x \frac{\mathrm{d}T^{}_{\nu}}{\mathrm{d}x} \left(\frac{\mathrm{d}\rho^{}_{\gamma}}{\mathrm{d}T^{}_{\gamma}}+\frac{\mathrm{d}\rho^{}_{e}}{\mathrm{d}T^{}_{\gamma}} \right) }{\frac{\mathrm{d}\rho^{}_{\gamma}}{\mathrm{d}T^{}_{\gamma}}+ \frac{\mathrm{d}\rho^{}_{e}}{\mathrm{d}T^{}_{\gamma}} } \;,
\end{align}
along with $m\, {\mathrm{d}z^{}_{i}}/{\mathrm{d}x} = x\, {\mathrm{d}T^{}_{i}}/{\mathrm{d}x} + T^{}_{i}$ for $i = \nu$ and $\gamma$. 
The temperature of the photon-electron plasma can also be derived by solving the electron collision terms similar to that of neutrinos. But since the energy conservation is a guaranteed result of microscopic processes, they are actually equivalent.
By combining Eqs.~(\ref{eqapp:boltz}), (\ref{eqapp:st}) and (\ref{eqapp:tgx}), we are ready to solve the background quantities for any given initial conditions. 

\end{appendix}



\bibliographystyle{utcaps_mod}
\bibliography{references}

\end{document}